# Photovoltaic effect by soft phonon excitation

Y. Okamura[1*,†], T. Morimoto[1,2†], N. Ogawa[1,3], Y. Kaneko[3], G-Y. Guo[4,5],
M. Nakamura[3], M. Kawasaki[1,3], N. Nagaosa[1,3], Y. Tokura[1,3,6] and Y. Takahashi[1,3*]

[1]*Department of Applied Physics and Quantum Phase Electronics Centre, University of Tokyo, Tokyo 113-8656, Japan*

[2] *PRESTO, Japan Science and Technology Agency (JST), Tokyo 113-8656, Japan*

[3]*RIKEN Centre for Emergent Matter Science (CEMS), Wako 351-0198, Japan*

[4]*Department of Physics and Centre for Theoretical Physics, National Taiwan University, Taipei 10617, Taiwan*

[5]*Physics Division, National Centre for Theoretical Sciences, Taipei 10617, Taiwan*

[6]*Tokyo College, University of Tokyo, Tokyo 113-8656, Japan*

[*]To whom correspondence should be addressed (okamura@ap.t.u-tokyo.ac.jp, youtarou-takahashi@ap.t.u-tokyo.ac.jp)

[†]These authors equally contribute to this work.

**Photodetection is an indispensable function of optoelectronic devices in modern communication and sensing systems[1]. Contrary to the near-infrared/visible regions, the fast and sensitive photodetectors operated at room temperature for the far-infrared/terahertz regions are not well developed despite a possibly vast range of applications[2-4]. The bulk photovoltaic effect (BPVE) in single-phase noncentrosymmetric materials based on the shift current mechanism enables less-dissipative energy conversion endowed with instantaneous responsivity owing to the quantum-mechanical geometric phase of electronic states[5-17]. Nevertheless, the small-band-gap material for the low-energy BPVE inevitably suffers from the thermal noise due to the intrinsically high conductivity. Here, we demonstrate the shift current induced by soft-phonon excitations without creation of electron-hole pairs in the archetypal ferroelectric $BaTiO_3$ by using the terahertz light, whose energy scale is three orders of magnitude smaller than the electronic band gap. At and above room temperature, we observe appreciable photocurrents caused by the soft-phonon excitation as large as that for electronic excitation and their strong phonon-mode dependence. The observed phonon-driven BPVE can be well accounted for by the shift current model considering the electron-phonon coupling in the displacement-type ferroelectrics as supported by the first-principles calculation. Our findings establish the efficient quantum BPVE arising from low-energy elementary excitations, suggesting the novel principle for the high-performance terahertz photodetectors.**

In the past decade, the essential role of the electronic topology characterized by the geometric phase (Berry phase) has been recognized for various classes of electromagnetic phenomena in crystalline solids[18]. In addition to the most established framework for the intrinsic anomalous Hall effect in the systems without time reversal symmetry[19], the recent extensive theoretical studies reveal the geometrical nature of nonlinear optical effects in the systems without inversion symmetry including polar or ferroelectric materials[5,20-23]. The shift current is one such mechanism of the BPVE arising from the quantum geometric phase and can be the promising principle for the solar cell and photodetector[9]. The interband optical transition induces the shift of the electronic wave packet in real space, i.e., shift vector composed of the Berry connections of valence and conduction bands involved in the optical transition, leading to the steady-state photocurrent without external bias. The BPVE by shift current is driven by the geometric nature of the Bloch wavefunctions and is essentially different from the transport current, producing the less-dissipative energy flow with almost instantaneous response. The shift current response is generated within the light illumination place and never dependent of the defect density of the material and therefore the mobility of the photocarrier[16,17]. This is contrasted to the widely used *p-n* junction, where the photocarriers travel to the electrodes via drift-diffusive transport induced by the built-in potential.

The modern theory of ferroelectricity is closely relevant to the intuitive picture of the shift current addressed in this work. The ferroelectric polarization is generally composed of the asymmetry of the valence electronic wavefunction in addition to the displacement of the ions. The former electronic polarization is formulated by the Berry phase of the valence bands, whose change upon the interband transition is responsible for the shift vector[23-25]. Recent experimental and theoretical studies have indeed demonstrated the large shift current response in various classes of materials hosting the large electronic polarization[6,7,9,11-13,16,17]. The direct connection between the electronic polarization and shift current indicates that the shift current generation can be driven by any excitation that is coupled with the electronic states and is able to modulate the electronic polarization, including low-energy elementary excitations[26,27]. Note, however, that the usual polarization current cannot be the direct current (DC) in sharp contrast to the shift current studied here. This prospective photocurrent mechanism does not require the electronic interband transition contrary to many other photovoltaic mechanisms, leading to the unprecedented low-energy photodetection, while its feasibility has never been examined.

Here we exploit the soft-phonon excitation in displacive-type ferroelectrics to realize this idea. The soft phonon has the same deformation pattern as the ferroelectric state; the softening of phonon frequency is accompanied by enhancement of oscillator strength, which is transferred from the electronic excitation. Therefore, the direct excitation of the soft phonons by the terahertz light modulates the electronic polarization as well, producing the shift current without creation of electron-hole pairs. On this basis, we targeted a ferroelectric $BaTiO_3$, which is known to host the sizable electronic polarization[28,29] and the strong soft-phonon excitations in terahertz region (< 15 meV)[30-32].

The paraelectric cubic structure of the perovskite-type $BaTiO_3$ turns into the ferroelectric tetragonal structure below $T_C$ ~ 390 K (Fig. 1a) with the spontaneous polarization along the *c* axis (Fig. 1b), and further successive transitions occur at ~ 280 K (into orthorhombic phase) and at ~ 190 K (into rhombohedral phase). In the tetragonal phase, two distinct phonon excitations associated with the ferroelectricity are observed in the terahertz spectra (Fig. 1c and 1d)[30-32]. The unfrozen optical phonon in tetragonal phase (Slater mode), which is associated with the displacement-type transition, has polarization perpendicular to the ferroelectric polarization (||*a*). In addition, reflecting the coexistence

of order-disorder-type transition nature, the fluctuation of arrangement of electric dipoles gives rise to the relaxational mode polarized along the *c* axis. BaTiO$_3$ is transparent below the large band gap energy ~ 3.2 eV except for the infrared optical phonon excitations (Fig. 1e), being consistent with its highly insulating nature. We note that the BPVE for the interband transition in ultraviolet region has been extensively studied in this material[6,33-37]. The photocurrent action spectra reported in the seminal work have been recently reproduced by the first-principles calculations[6], suggesting the dominant role of the shift current mechanism for the BPVE.

To pursue the phonon-driven shift current, we measured the photocurrent induced by the terahertz pulse focused on the centre of the sample, away from the metal contacts (Fig. 1f and 2a; see also Methods). We note that the shift current is generated only within the illumination place but can flow into the electrodes via the capacitive coupling in case of the pulse excitation[12,16]. The linearly polarized terahertz light, whose centre photon energy and bandwidth are both ~ 4 meV (Fig. 1f), is suitable for the resonant excitation of two types of soft phonons. The BPVE is generally described by the second-order nonlinear optical conductivity tensor $\sigma^{(2)}$. On the basis of the symmetry consideration, the *zzz* ($\sigma^{(2)}_{zzz}$) and *zxx* ($\sigma^{(2)}_{zxx}$) components of the $\sigma^{(2)}$ are allowed in the tetragonal phase (see Fig. 1a for the coordinates); here we define $j_i = \sigma^{(2)}_{ijk}(\omega)E_jE_k$, where $j_i$ and $E_j$ represent the photocurrent density along the *i* axis and electric field of light along the *j* axis, respectively. Accordingly, the $\sigma^{(2)}_{zzz}$ and $\sigma^{(2)}_{zxx}$ describe the photocurrent responses upon the relaxational mode for *E*∥*z*(*c*) and upon the Slater mode for *E*∥*x*(*a*), respectively.

For the both cases, we observed appreciable zero-bias photocurrent pulses in the single-domain ferroelectric states at room temperature (Fig. 2b and 2c). The signs of the photocurrents are totally reversed by the reversal of the ferroelectric polarization (red and blue curves in Fig. 2b; magenta and light blue curves in Fig. 2c). These signals are gradually enhanced with increasing the temperature and disappear in the centrosymmetric paraelectric phase (Fig. 2d), while the strong soft phonon excitations exist in terahertz region even above the $T_C$[30,32]. We note that the photocurrent signal is least observed for the multi-domain state with no net polarization (*P*) after zero field cooling (gray curve in Fig. 2b). Therefore, it is concluded that the observation of photocurrents is produced by the inversion symmetry breaking of the bulk crystal. The magnitude of the photocurrent is proportional to the square of the terahertz electric field (Fig. 2e), in accord with the second-order nonlinear effect described by $\sigma^{(2)}_{ijk}$. We also found that the light-polarization dependence of the photocurrent shows the sine-wave-like behaviour (Extended Data Fig. 1), which confirms the tensorial nature of $\sigma^{(2)}$ and is consistent with the point group of BaTiO$_3$ at room temperature (4*mm*). All these results manifest the presence of the BPVE without interband transition, i.e., the phonon-driven terahertz shift current.

This scenario is corroborated by the following two observations: First, as seen in Fig. 2f, we observed the sharp photocurrent pulse only during the irradiation of the terahertz-light pulse. The time integral of the photocurrent shows the step-function-like behaviour, which persists without discernible decay or increase (orange curve in Fig. 2f). This observation is consistent with the fact that the shift current is the steady-state photocurrent during the light irradiation with nonvanishing integrated current. This is in contrast to the transient pyroelectric current induced by the instantaneous light-induced heating with zero integrated current in total (for more details of the transient pyroelectric current, see Extended Data Fig. 2 and Methods). The other important observation is that the photocurrent signal is independent of the external bias voltage (Fig. 2g), indicating that the open circuit voltage is larger than 600 V/cm. Since the open circuit voltage is at most 80 V/cm for the above-band-gap excitation[34], this large open circuit voltage demonstrates the absence of mobile photocarriers generated by the terahertz field. Thus, the diffusive charge transport in the photocurrent generation can be excluded for this terahertz photocurrent generation, as expected for the phonon-driven shift current.

The clear phonon-mode dependence is observed in the nonlinear optical conductivity, which is directly connected to the shift vector[20]. We note that the duration of the original photocurrent pulse induced inside the material should be comparable to that of the irradiated terahertz pulse (~ 1 ps) due to the instantaneous response expected for the shift current mechanism. However, the electrical circuit used for the ultrafast photocurrent detection inevitably causes the attenuation and broadening of current pulses to ~ 5 ns, making the quantitative estimation of $\sigma^{(2)}$ difficult. Thus, here we introduce the alternative quantity $\alpha_j J^P_i/I_{abs}$, which is proportional to the nonlinear optical conductivity $\sigma^{(2)}_{ijj}$ (for more details, see Methods); $\alpha_j$, $J^P_i$ and $I_{abs}$ represent the absorption coefficient for the light polarization along the $j$ axis, the peak value of pulsed photocurrent flowing along the $i$ axis $J_i(t)$, and the absorbed light power, respectively. In terahertz region, the $zxx$ component (Slater mode, $\alpha_x J^P_z/I_{abs}$) is ~ 24 times larger than the $zzz$ one (relaxational mode, $\alpha_z J^P_z/I_{abs}$) (Fig. 3a), demonstrating the strong phonon-mode dependence of the $\sigma^{(2)}$. In contrast, the electronic excitations show the relatively isotropic $\sigma^{(2)}$ spectra, whose magnitudes are comparable to that of the $zxx$ component (Slater mode) for the phonon excitation (Fig. 3b; see also Extended Data Fig. 3). Accordingly, the Slater mode gives rise to the shift current as large as the interband optical transitions.

We also evaluated the Glass coefficient in a similar manner[38]; it represents the photocurrent divided by the absorbed light power, which can be quantified by $J^P_i/I_{abs}$ (see also Methods). On the basis of Extended Data Fig. 4, the Glass coefficients for the two soft phonon modes are found to be comparable to those for the electronic excitations in BaTiO$_3$. The Glass coefficient of this material in the ultraviolet region is known to be relatively large among other known compounds[14], indicating the high efficiency of the phonon-driven shift current.

To understand the mechanism of phonon-driven shift current, we develop the theoretical framework (for more details, see Supplementary Note 1). We consider the Rice-Mele model, which is a famous one-dimensional model for ferroelectrics[23]. By taking into account the electron-phonon coupling with strength of $g$, we derive the nonlinear optical conductivity $\sigma^{(2)}$ for the shift current upon the phonon excitation with resonance energy $E_{ph}$ using the diagrammatic technique[39]. The magnitude of $\sigma^{(2)}$ is scaled by square of the strength of the electron-phonon coupling $g^2$ and inverse phonon energy $1/E_{ph}$; $\sigma^{(2)} \propto g^2/E_{ph}$ (see Eq. (S2) of Supplementary Note 1 for explicit expression). Therefore, the intrinsically small energy of soft phonon should enhance the $\sigma^{(2)}$ upon the phonon excitation, rationalizing the large phonon-driven shift current comparable to the electronic shift current. This minimal theoretical analysis also explains the phonon-mode dependence of $\sigma^{(2)}$. The nonlinear conductivity can be rewritten as, $\sigma^{(2)} = \frac{e}{E_{ph}}\sigma^{(1)}R$ with using the linear optical conductivity $\sigma^{(1)}$ and the shift vector $R$ representing real-space shift of the electronic wavefunction for the phonon excitation. The magnitude of $\sigma^{(1)}$ for the Slater mode is ~ 70 times larger than that for the relaxational mode at ~ 4 meV. If these modes have comparable shift vector $R$, this accounts for the strong anisotropy of $\sigma^{(2)}$ as experimentally observed.

On the basis of the formalism established here, we calculate the $\sigma^{(2)}_{zxx}$ due to the Slater mode using the density functional theory (for the details, see Supplementary Note 2). The magnitude of the $\sigma^{(2)}_{zxx}$ due to the Slater mode is comparable to or larger than that of the electronic excitation near the band gap and the sign of the $\sigma^{(2)}_{zxx}$ is opposite to that of the electric polarization direction (Fig. 4). These facts reasonably explain the experimental observation, which corroborates the shift current origin of terahertz photocurrent.

Photodetection of the terahertz/far-infrared light is relevant to a variety of applications including sensing, imaging and communication. The conventional low-energy photodetection relies either on the photocarrier generation in the small-band-gap materials or on the photo-induced thermalization of the systems[2], which usually cannot ensure both high-speed and low-noise responses at room temperature simultaneously. In this context, the phonon-driven shift current can realize both almost instantaneous response and significant reduction of the thermal noise due to the absence of conduction carriers even at room temperature. Our present work proves the phonon-driven shift current without creation of real electron-hole pairs, which demonstrates the crucial importance of the geometrical nature of inversion-broken materials for the efficient photocurrent generation. The phonon excitation may be strongly coupled to the topological electronic structure in particular materials, which will open up a new avenue to control the topological nature of matter[40,41]. These findings lead to the novel principle for the high-performance low-energy photodetectors operated irrespective of the band gap and even stimulate the further research for hitherto unexplored topological quantum phenomena with promising functionality.

**Acknowledgement**

We thank S. Sawamura, M. Ogino and J. Fujioka for experimental help. G.-Y.G thanks Samual Poncé for helpful communications on the EPW program. This work was partially supported by JST CREST (JPMJCR16F1 and JPMJCR1874). N.N. is supported by JSPS KAKENHI Grant numbers 18H03676. Y. Takahashi is supported by JSPS KAKENHI Grant numbers 21H01796. G.-Y.G is supported by the Ministry of Science and Technology, the National Center for High-performance Computing and the Center for Emergent Materials and Advanced Devices, National Taiwan University, Taiwan.

**Author contributions**

N.N., Y. Tokura and Y. Takahashi conceived the project. Y.O., N.O. and Y. Takahashi performed the optical measurement and analysed the data with assistance of M.N. and M.K. Y.K. grew the single crystal. T.M. and N.N. developed the theory. G.-Y.G. performed the first-principles calculation. Y.O., T.M., N.O., N.N., Y. Tokura, and Y. Takahashi discussed and interpreted the results with inputs from other authors. Y.O., T.M. and Y. Takahashi wrote the manuscript with assistance of other authors.

**Additional information**

**Competing interests:** The authors declare no competing interests.

**Methods**

**Single crystal growth.** The single crystalline samples of Ba$_{0.97}$Sr$_{0.03}$TiO$_3$ were grown by the laser floating zone method[42]. Here, Sr is dilutely doped for a technical reason; a large-size single crystal can be acquired by avoiding the incorporation of hexagonal (non-perovskite) crystal structural phase during the crystal growth.

**Electric polarization measurement.** The electric polarization is deduced by measuring the pyroelectric current with increasing the temperature. The single-domain state is stabilized by the field cooling from ~ 410 K well above $T_C$ with application of the electric field of ±2.85 kV/cm.

**Intense terahertz pulse generation.** For a light source, we used a regenerative amplified Ti:sapphire laser system with the pulse energy of 5 mJ, the pulse duration of 100 fs, the repetition rate of 1 kHz, and the centre wavelength of 800 nm. The intense terahertz pulses were generated by the tilted-pulse front method with a $LiNbO_3$ crystal[43]. The spot size was about 1 mm. The time waveform of the terahertz pulse is measured by the electro-optic (EO) sampling technique with using a ZnTe (110) crystal. The intensity of the terahertz light is controlled by using the wire grid polarizers. Except for investigating the terahertz-electric-field dependence (Fig. 2e), we use the terahertz pulse with the maximal intensity shown in Fig. 1f. The terahertz power is estimated by using a power meter (T-RAD, GENTEC-EO).

**Ultraviolet light generation.** We used the second harmonic generation of the near-infrared/visible light from an optical parametric amplifier pumped by a Yb-based amplified laser with the pulse energy of 0.833 mJ, the pulse duration of ~ 140 fs and the repetition rate of 6 kHz. The spot size was about 1 mm.

**Pulsed photocurrent measurement.** We measured the pulsed photocurrent signal for each light pulse by using an oscilloscope. The bandwidth of the preamplifier was 200 MHz. The sample size is typically 2.5x3x1 $mm^3$. We shined the light at the centre of the sample and carefully covered the electrodes with the aluminium foil to eliminate the spurious effects. The photocurrent pulse duration is determined presumably by the bandwidth of the preamplifier (200 MHz), as indicated by the duration of the main photocurrent pulse (~ 5 ns). Meanwhile, we also observe the similar-shaped secondary and even tertiary pulse signals, leading to more broadened pulse. These parasitic signals are known as the ringing effect, which inevitably occurs for a short electrical pulse due to the parasitic capacitances and inductances (and impedance mismatch) in the circuit. We note that this oscillatory parasitic signal is not relevant to the piezoelectric effect induced by the coherent phonon excitation because the same waveform of transient photocurrent is observed for UV pulse excitations (Extended Data Fig. 3).

**Optical conductivity and absorption spectra.** Polarized optical conductivity and absorption spectra above 0.07 eV were obtained through the Kramers-Kronig analysis of the reflectivity spectra. We measured the reflectivity spectra using a Fourier-transform-type spectrometer in the infrared region above 0.07 eV and a monochromator-type spectrometer in the visible and ultraviolet regions below 4.1 eV. The reflectivity data above 4.1 eV were taken from ref. 44. $\sigma_{aa}$ and $\sigma_{cc}$ spectra below 0.07 eV were calculated from the dielectric constants shown in ref. 31 and ref. 32, respectively.

**Contribution from the photo-thermal effect.** The transient light-induced heating changes the ferroelectric polarization, which leads to the pyroelectric current. This photo-thermal effect might contribute to the apparent photocurrent. However, the presently observed photocurrent via the soft phonon excitations is verified to involve least pyroelectric-current component, as we discuss below.

We perform a comparative experiment to evaluate the pyroelectric current directly. We fabricate the 100-nm-thick AuPd film on top of the single-ferroelectric-domain BaTiO$_3$ sample (Extended Data Fig. 2a and 2b). The film almost totally absorbs the irradiated light except for the reflection loss ~ 50 % (transmittance is less than 0.2 %), therefore the light pulse can be regarded as the heat pulse for the sample. In Extended

Data Fig. 2c, we compare the pulsed photocurrents divided by the absorbed light power, $J/I_{\text{abs}}$, for the BaTiO$_3$ without and with the AuPd film under the 400-nm pulse irradiation. We also show the absorbed power dependence of these photocurrent responses in Extended Data Fig. 2d. Therefore, the photovoltaic effect for the BaTiO$_3$ with the AuPd film is negligible compared with that for bare BaTiO$_3$, which demonstrates that the photo-thermal effect can be neglected in the present experiment. In addition, we measure the temperature dependence of the photocurrent for the AuPd coated crystal, which shows a steep enhancement near the transition temperature as expected for the pyroelectric current (Extended Data Fig. 2e). Contrary to this, the observed photocurrent via the soft phonon excitations shows a much (more than 25 times) larger magnitude (Extended Data Fig. 2d) and a moderate temperature dependence toward $T_C$ (Fig. 2d) than the pyroelectric current, corroborating its shift current origin with the least photo-thermal pyroelectric-current component.

Finally, we estimate the possible photothermal effect owing to the light-induced heating. The increase of the sample temperature $\Delta T$ for the illumination place is estimated to be $I_{\text{abs}}/(CV\rho) = 9.5 \times 10^{-3}$ K, where $C$, $I_{\text{abs}}$, $V$ and $\rho$ are the heat capacity, absorbed power, sample volume where the light is absorbed and density of the crystal, respectively. The integrated pyroelectric current is calculated as $\Delta T \times |dP/dT| \times S = 0.033$ pC, where $S$ represents the cross section of excitation area. Note that here we omit the diffusion of excited phonon and heat, which substantially reduces the pyroelectric current, so that this estimation gives the upper limit of pyroelectric current. In fact, the experimentally observed pyroelectric current (Extended Data Fig. 2d) is as large as 20 % of this estimated upper limit. Both upper limit and observed pyroelectric current cannot account for the phonon-induced photocurrent ~ 0.15 pC (Fig. 2f). Therefore, we can conclude that the most part of photocurrent arises from the BPVE of soft phonon excitation.

**Evaluation of the nonlinear optical conductivity and Glass coefficient.** The photocurrent response can be characterized by the second-order nonlinear optical conductivity $\sigma^{(2)}_{ijj}$ and Glass coefficient $G_{ijj}$. The total current $J_i$ is described for the incident light normal to the sample surface by[6],

$$J_i = \frac{\sigma^{(2)}_{ijj}}{\alpha_j} E_j^2 w = G_{ijj} i_{\text{abs}} w,$$

where $\alpha_j$ is the absorption coefficient polarized along the $j$ axis, $E_j$ is the electric field of light along the $j$ axis, $i_{\text{abs}}$ is the absorbed power density of the light, and $w$ is the width of the sample surface. Since the original ultrafast shift current pulse induced inside the material should be much attenuated through the electrical circuit given the limited bandwidth of the preamplifier (200 MHz), we discussed the $\alpha_j J^P_i/I_{\text{abs}}$ and $J^P_i/I_{\text{abs}}$, which can quantify $\sigma^{(2)}_{ijj}$ and $G_{ijj}$, respectively. $J^P_i$ and $I_{\text{abs}}$ denote the peak value of the photocurrent pulse flowing along the $i$ axis and the absorbed light power, respectively. We note that, because we used the same sample width $w$ and the same spot size of the light, we can discuss in terms of $I_{\text{abs}}$ as well as $i_{\text{abs}}$. In addition, the light is totally absorbed for the present thick sample (thickness ~ 1 mm) except for the reflection loss, and therefore, the absorbed light power $I_{\text{abs}}$ can be calculated from $I(1-R)$, where $I$ and $R$ denote the irradiated light power and the reflectivity of the sample, respectively.

**Data availability.** The data that support the plots of this study are available from the corresponding author upon reasonable request.

**Figures**

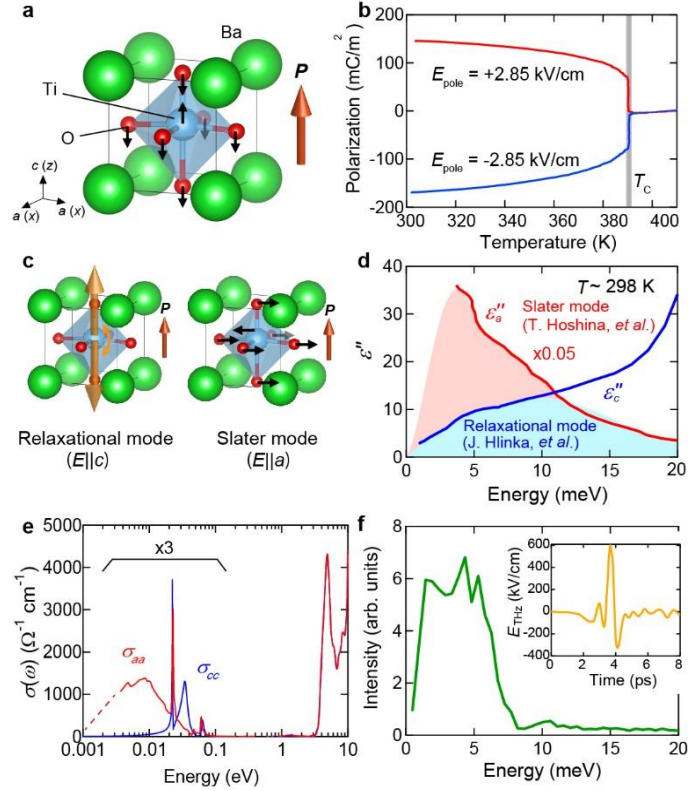

**Fig. 1| Static and dynamical ferroelectric nature of tetragonal BaTiO$_3$. a,** The crystal structure of the tetragonal phase of BaTiO$_3$. **b,** Temperature dependence of the spontaneous electric polarization along the $c$ axis. $E_{\text{pole}}$ represents the poling field. **c,** Schematic illustrations of two kinds of soft phonon modes; relaxational mode for $E\|c$ (left) and Slater mode for $E\|a$ (right). **d,** The imaginary part of the terahertz dielectric constants, $\varepsilon_a''$ (red curve) and $\varepsilon_c''$ (blue curve) in the tetragonal phase at room temperature. $\varepsilon_a''$ and $\varepsilon_c''$ spectra are reproduced from refs. 31 and 32, respectively. $\varepsilon_a''$ is multiplied by a factor of 0.05 for clarity. **e,** The optical conductivity spectra, $\sigma_{aa}$ (red curve) and $\sigma_{cc}$ (blue curve), in the wide energy range. The data below 0.1 eV are multiplied by a factor of 3 for clarity. **f,** The spectral amplitude of the terahertz electric field used in this study. (Inset) Time waveform of the terahertz light pulse.

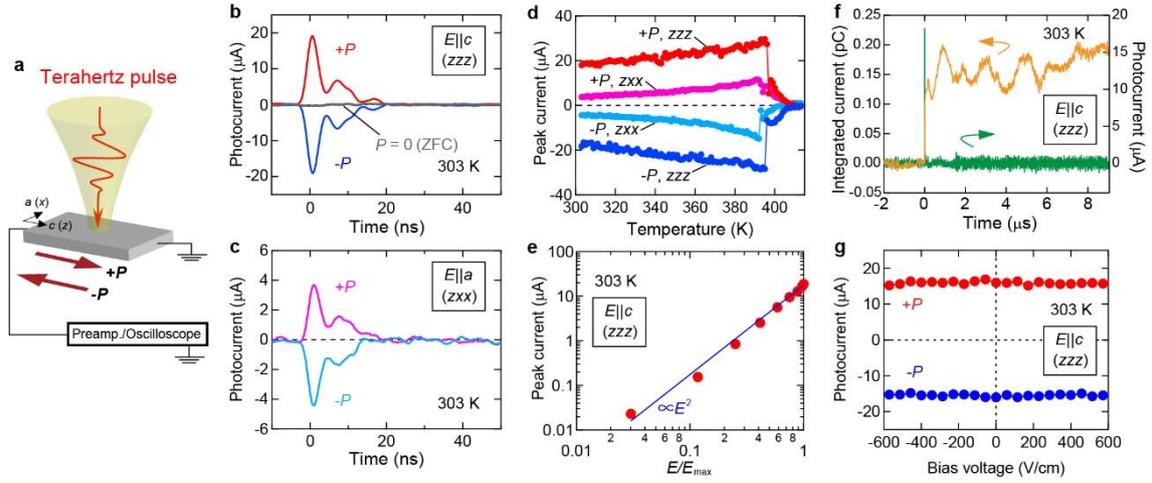

**Fig. 2| Terahertz bulk photovoltaic effect. a,** Schematic illustration of the experimental setup. **b,c,** Pulsed photocurrent for $E\|c$ (**b**) and $E\|a$ (**c**). In **b** (**c**), the red (magenta) and blue (light blue) solid curves represent the data at 303 K for $+P$ and $-P$ states, respectively. Photocurrent responses for $E\|c$ and $E\|a$ correspond to the $zzz$ and $zxx$ components of nonlinear optical conductivity, respectively. The gray curve in **b** represents the photocurrent for the multi-domain state after zero field cooling (ZFC). **d,** Temperature dependence of the peak values of photocurrents for the Slater mode and the relaxational mode. Red (magenta) and blue (light blue) circles denote the $zzz$ ($zxx$) components of nonlinear optical conductivity for $+P$ and $-P$ states, respectively. **e,** Terahertz electric field dependence of the peak value of the photocurrent (red circles) in the log-log scale. The blue line represents the fitting curve using $E^2$. **f,** Transient photocurrent response on the long time scale (green curve) and the corresponding integrated current (orange curve). **g,** The bias-voltage dependence of the peak values of the photocurrent pulses for $+P$ (red circles) and $-P$ (blue circles) states.

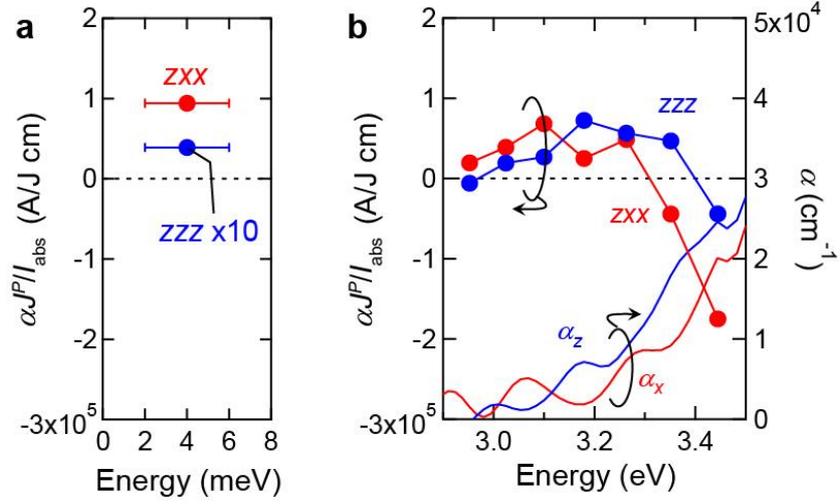

**Fig. 3| Nonlinear optical conductivity spectra for phonon and electronic excitations.**
**a,b,** *zxx* (red circles) and *zzz* (blue circles) components of $\alpha J^P/I_{abs}$, which respectively quantify the nonlinear optical conductivities $\sigma^{(2)}_{zxx}$ and $\sigma^{(2)}_{zzz}$, in terahertz (**a**) and visible/ultraviolet regions (**b**). $\alpha$, $J^P$, and $I_{abs}$ represent the absorption coefficient, peak value of pulsed photocurrent and absorbed light power, respectively. In calculating the terahertz $\alpha J^P/I_{abs}$, we used the $\alpha$ at 4 meV obtained from refs. 31 and 32. The horizontal bars shown in the panel **a** denote the bandwidth of the terahertz light. On the right axis of the panel **b**, we show the absorption spectra $\alpha_x$ (red curve) and $\alpha_z$ (blue curve).

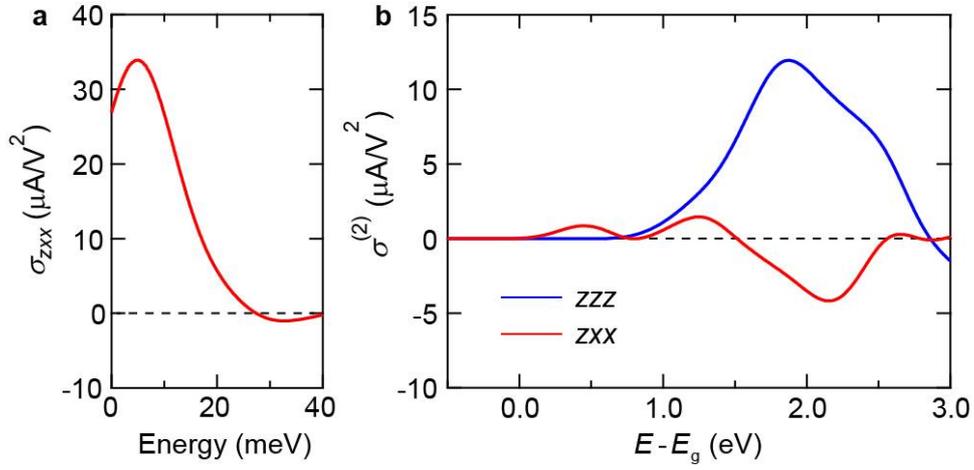

**Fig. 4| The shift current conductivity from first-principles calculations. a,** Calculated shift current response due to the phonon excitation with broadening $\Gamma = 10$ meV. **b,** Calculated shift current response due to the electronic excitation with carrier life-time broadening $\Gamma = 0.1$ eV. $E$ and $E_g$ shown in the horizontal axis of **b** represent the photon energy and band gap, respectively. Here we define the photocurrent opposite to the electric-polarization direction as the positive sign.

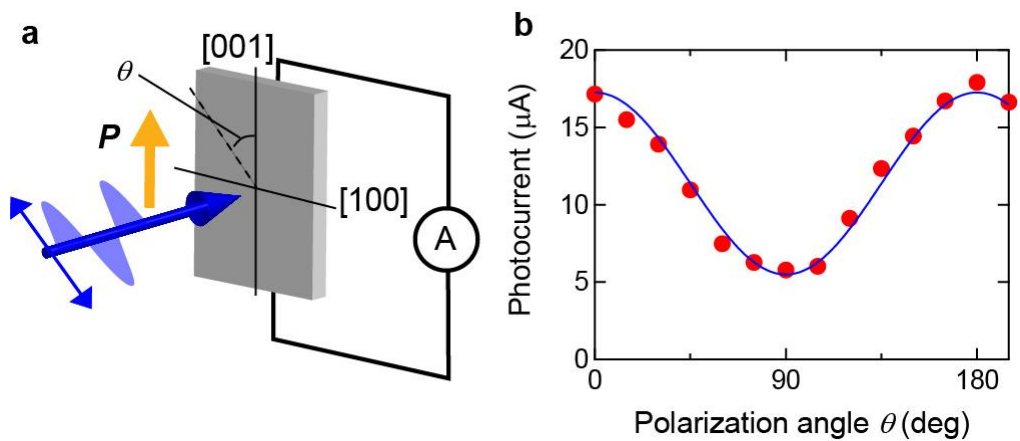

**Extended Data Fig. 1| Light-polarization dependence of the photocurrent. a,** Schematic illustration of the experimental setup. The $\theta$ represents the angle between the terahertz electric field and [001] axis. The photocurrent along [001] axis is measured. **b,** Polarization dependence of the photocurrent (red circles). The blue curve represents the fitting curve using $j_0 + j_1 \cos 2\theta$.

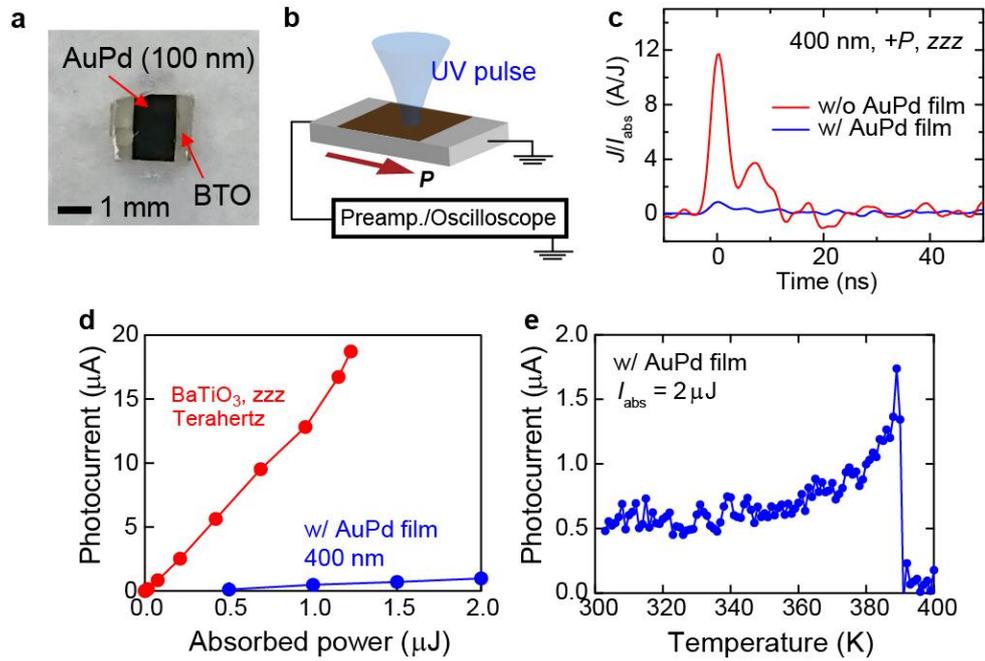

**Extended Data Fig. 2| Contribution from photothermal effect. a,** The picture of the sample. We fabricated the 100-nm-thick AuPd film on top of the single-ferroelectric-domain BaTiO$_3$. **b,** Schematic illustration of experimental setup. **c,** Time traces of $J/I_{abs}$ for single-domain BaTiO$_3$ at room temperature without (red curve) and with (blue curve) AuPd films, where $J$ represents the pulsed photocurrent. We measured the *zzz* component under the 400-nm pulse irradiation. **d,** The absorbed power dependence of the photocurrent for uncoated BaTiO$_3$ crystal with the terahertz excitation (red circles) and AuPd coated one (blue circles). **e,** Temperature dependence of the photocurrent for the coated crystal.

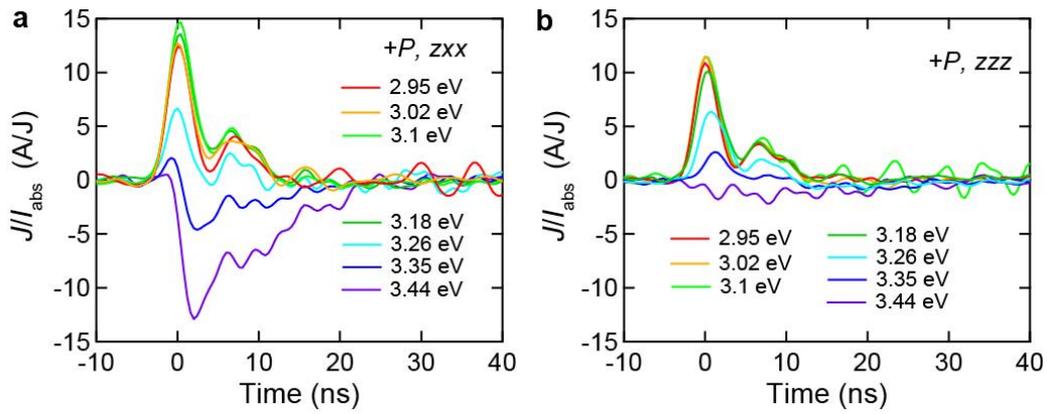

**Extended Data Fig. 3| Photocurrent response for the electronic excitation. a,b,** The time trace for the *zxx* (**a**) and *zzz* (**b**) components of $J/I_{abs}$ for each excitation energy in the single ferroelectric domain (+*P*) state.

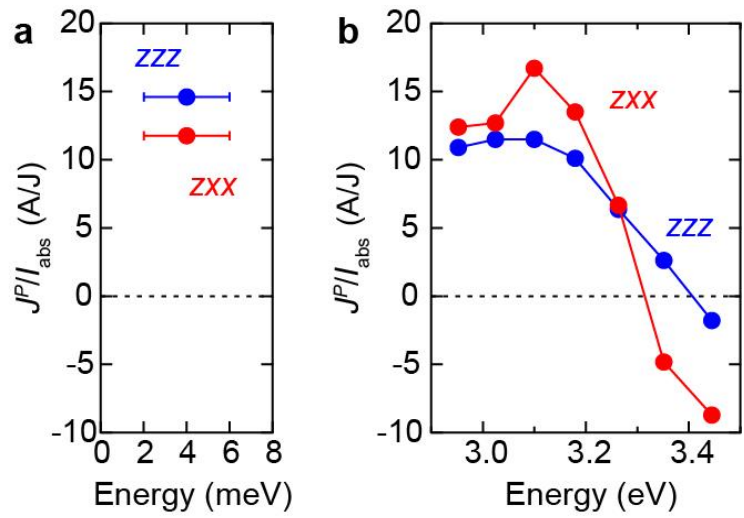

**Extended Data Fig. 4| Glass coefficient for phonon and electronic excitations. a,b,** The *zxx* (red circles) and *zzz* (blue circles) components of $J^P/I_{abs}$, which respectively quantify the Glass coefficients $G_{zxx}$ and $G_{zzz}$, in terahertz (**a**) and visible/ultraviolet regions (**b**). The horizontal bars shown in the panel **a** denote the bandwidth of the terahertz light.